\documentclass[final,authoryear,5p,times,twocolumn]{elsarticle}

\usepackage{graphicx}

\usepackage{amssymb}

\usepackage[pdftex,pdfpagemode={UseOutlines},bookmarks,bookmarksopen,colorlinks,linkcolor={blue},citecolor={green},urlcolor={red}]{hyperref}
\usepackage{hypernat}

\usepackage{listings}
\usepackage{color}

\definecolor{mygreen}{rgb}{0,0.5,0}
\journal{Astronomy \& Computing}

\usepackage{upquote}

\begin{document}

\begin{frontmatter}



\title{AST: A library for modelling and manipulating coordinate systems\tnoteref{ascl}}
\tnotetext[ascl]{This code is registered at the ASCL with the code
  entry ascl: \href{http://www.ascl.net/1404.016}{1404.016}}


\author[jac,eao]{David S.\ Berry\corref{cor1}}
\ead{d.berry@eaobservatory.org}
\author[ral]{Rodney F.\ Warren-Smith}
\author[jac,lsst]{Tim Jenness}

\cortext[cor1]{Corresponding author}

\address[jac]{Joint Astronomy Centre, 660 N.\ A`oh\=ok\=u Place, Hilo, HI
  96720, USA}
\address[eao]{East Asian Observatory, 660 N.\ A`oh\=ok\=u Place, Hilo, HI
  96720, USA}
\address[ral]{RAL Space, STFC Rutherford Appleton Laboratory, Harwell Oxford, Didcot, Oxfordshire OX11 0QX, UK}
\address[lsst]{LSST Project Office, 933 N.\ Cherry Ave, Tucson, AZ~85721, USA}

\begin{abstract}
In view of increased interest in object-oriented systems for describing
coordinate information, we present a description of the data model used
by the Starlink AST library. AST provides a comprehensive range of
facilities for attaching world co-ordinate systems to astronomical data,
and for retrieving and interpreting that information in a variety of
formats, including FITS-WCS. AST is a mature system that has been in use
for more than 17 years, and may consequently be useful as a means of
informing development of similar systems in the future.
\end{abstract}


\end{frontmatter}


\newcommand{\mnras}{Mon Not R Astron Soc}
\newcommand{\aap}{Astron Astrophys}
\newcommand{\aaps}{Astron Astrophys Supp}
\newcommand{\pasp}{Pub Astron Soc Pacific}
\newcommand{\apj}{Astrophys J}
\newcommand{\apjs}{Astrophys J Supp}
\newcommand{\qjras}{Quart J R Astron Soc}
\newcommand{\an}{Astron.\ Nach.}
\newcommand{\ijimw}{Int.\ J.\ Infrared \& Millimeter Waves}
\newcommand{\procspie}{Proc.\ SPIE}
\newcommand{\aspconf}{ASP Conf. Ser.}

\newcommand{\ascl}[1]{\href{http://www.ascl.net/#1}{ascl:#1}}

\section{Introduction}
\label{sec:intro}

The Starlink AST library \citep[][\ascl{1404.016}]{SUN211} provides a
generalised scheme for modelling, manipulating and storing inter-related
coordinate systems. Whilst written in C, it has bindings for several
other languages including Python, Java, Perl and Fortran. It has
specialised support for many of the coordinate systems and projections
commonly used to describe astronomical World Coordinate Systems (WCS),
including all the celestial and spectral coordinate systems described by
the FITS-WCS standard \citep{FITSWCSI,FITSWCSII,FITSWCSIII}, plus various
popular distortion schemes currently in use. However, it is not limited to
WCS, and may be used in any situation requiring transformation between
different coordinate systems.

Unlike FITS-WCS, which supports only a relatively small set of prescribed
transformation recipes reflecting the coordinate transformations within
an optical telescope, AST allows arbitrarily complex transformations to be
constructed by combining simple atomic transformations. This allows a much
wider range of transformations to be
described than is possible using FITS-WCS, and so can accommodate a wider
range of data storage forms without the need to re-grid the data.

AST was released in 1998 \cite[][included in ``Twenty Years of
ADASS''; \citet{adass20}]{1998ASPC..145...41W}. Since then it has been in
continuous use within the Starlink Software Collection \citep[][\ascl{1110.012}]{2014ASPC..485..391C}
and is also used by various other major astronomical software tools such as
DS9 \citep[][\ascl{0003.002}]{2003ASPC..295..489J} and SPLAT-VO \citep[][\ascl{1402.008}]{2014A&C.....7..108S}.

Interest in flexible schemes for representing inter-related coordinate
systems has increased recently --- for instance, in the discussions
about possible successors to the FITS format
\citep{2015Mink,2015A&C....12..240G,2016Shortridge}, and within the
Astropy \citep{2013A&A...558A..33A,P028_adassxxv} and International
Virtual Observatory Alliance (IVOA) (STC2; Rots, in preparation)
projects. These discussions suggest it is an appropriate time to
review the lessons learned from AST.

This paper first presents an account of the historical issues that drove
the initial development of AST, together with the reasoning behind some
of the design decisions, and then presents an over-view of the more
important aspects of the data model used by AST.

\section{Historical Perspective}

\subsection{Initial Problems}

The first public release of the AST library was in 1998
\citep{1998StarB..20....6L,1998StarB..20....7D} but some of the
underlying concepts date from the late 1980s, when the Starlink
Project was designing its NDF data format for gridded astronomical
data \citep[see][and references therein]{2015Jenness}. There was clearly a need to relate positions
within gridded data, using coordinates based on pixel indices, to
real-world positions on the sky, wavelengths in a spectrum and so
on.

Calibration of spectra, for example, was commonly performed by fitting
a polynomial to express wavelength as a function of pixel position and
then either storing the polynomial coefficients, or tabulating the
polynomial value at each pixel centre. While not completely general,
the latter option was an acceptable solution and was adopted as part
of the Starlink data format. An array giving the central wavelength at
each pixel was stored as the \texttt{AXIS} component in the NDF data
structure and did good service in spectroscopic applications. It was
also possible, in a simple minded way, to attach an \texttt{AXIS}
array to each dimension of an image or any gridded data set of higher
dimensionality. This allowed each of its axes to be calibrated in
terms of world coordinates.

This approach was adequate if the axes represented independent
quantities (like wavelength and position for a long-slit spectrum),
but did not suffice if the axes were inter-independent. Unfortunately,
in the common case of celestial coordinates (such as Right Ascension
and Declination), the axes are almost always inter-dependent. This is
because the sky is essentially spherical and its coordinates are
therefore naturally curvilinear when projected into two
dimensions. This inter-dependence is a common feature of world
coordinate systems in practice, so a solution was clearly needed that
addressed it properly.

The Flexible Image Transport System
\citep[FITS;][]{1981A&AS...44..363W,1995ASPC...77..233G}, at that
time, addressed the issue in a better but still rudimentary way. In
essence, it stored a physical pixel size (\emph{e.g.}, in seconds of arc),
allowed for a linear scaling of an image (typically to allow for the
position angle rotation of the telescope) and then projected it on to
the celestial sphere using one of a defined set of map
projections. This representation was clearly based on a model of a
physical telescope and how it imaged an observed region of the sky in
its focal plane.

While successfully accommodating the curvilinear nature of sky
coordinates, this FITS approach was still limited in many ways. In
essence, it defined a small set of functional forms (based on map
projections) through which pixel coordinates could be mapped on to
celestial coordinates and back again. However, if the actual
relationship between pixel coordinates and world coordinates didn't
correspond to one of these functional forms, then it wasn't possible
to use FITS to store the coordinate information\footnote{Unless the data was
first re-gridded into a form supported by FITS.}.

For instance, if astronomical instrumentation were to use a novel map
projection, if arbitrary instrumental distortions were present or if
the data were re-gridded into a non-physical space, then the FITS
approach would fail. It also had limited support for high-accuracy
astrometry, where the departure of the sky from a perfect sphere, for
a variety of reasons, has to be taken into account. In addition, there
are many other non-celestial world coordinate systems that one might
use (involving energy, velocity, time, frequency, \emph{etc}.) that no
contemporary system could represent adequately.

Unfortunately, this list of limitations only scratches the surface of
the problem as it was perceived at the time. Other considerations,
such as the time-dependent relationship between non-inertial celestial
coordinate systems, the dependence of apparent positions on the
position and velocity of the observer (and also on the wavelength of
observation and atmospheric conditions) and periodic revisions to the
fundamental definitions of celestial coordinate and time systems would
all have to be accommodated, as would numerous other issues specific
to particular domains (celestial coordinates, time systems, radial
velocities, wavelength/energy, \emph{etc}.). This was several years before the
FITS community commenced work on what was eventually to become the current
FITS-WCS standard.

\subsection{TRANSFORM}

In the late 1980s, no immediate and general solution to these problems
could be seen. Recognising the limitations in the FITS approach,
however, the Starlink Project decided to take a hard line and to omit
completely any component dedicated to world coordinate systems from
its new NDF data format. Instead, this \emph{astrometry extension}
(from which the name AST is derived) was to be added at a later date
when a suitable solution had been formulated.

This decision was undoubtedly strongly influenced by Patrick Wallace's
presence in the Project and the major work he had done on the SLALIB
library \citep[][\ascl{1403.025}]{1994ASPC...61..481W} to encapsulate best-practice in
astrometric calculations (and also in other domains such as time
systems). Discussion within the Project rapidly convinced us that if
we adopted the FITS approach as it existed at the time, we would cut
ourselves off from the proper rigorous treatment of astrometric data
that is needed for the highest accuracy.

Consequently, a pilot project was conducted to explore alternative
approaches. The most important limitation of the FITS approach was felt
to be the use of a fixed set of functional forms (map projections) each
of which was associated with a small fixed set of parameters. This
simplified storing the information in 80-character FITS \emph{header cards},
but clearly the set of functional forms that might ultimately be needed
was much larger than had been recognised. Adding new ones might become a
never-ending project and that, in turn, raised the prospect of
continually upgrading all software that had to read and process FITS
headers and handle coordinate systems.

The alternative approach that we explored was to write an expression
parser that would accept sets of arithmetic expressions similar to those
used in Fortran and C, along with the usual set of mathematical
functions. Together with a method of passing named parameter values into
these expressions, this greatly increased the set of functional forms
that could be represented. The expressions themselves (encoded as
character strings) and the associated parameter values could easily be
stored in astronomical data sets. Typically, one set of expressions would
relate pixel coordinates to world coordinates (\emph{e.g.}, sky coordinates) and
a second, optional, set would define the inverse transformation. The
expression syntax was powerful enough to represent a wide range of map
projections plus many other transformations into alternative world
coordinate systems.

A processing engine was also provided that could use the stored
expression data to transform actual coordinate values.

A library implementing this, called \textsc{transform}, was released
in 1989 \citep{SUN61,1989StarB...4....7L}. It stored its data (the
expressions and parameters) in Starlink's Hierarchical Data Format
\citep[HDS;][]{SUN92,SSN27,2015HDS} and was thus able to integrate with the
Starlink NDF data format to attach arbitrary world coordinates to
gridded astronomical data sets.

\subsection{TRANSFORM Lessons}

Ultimately, \textsc{transform} turned out not to be a full solution to
the WCS problem and did not become part of the NDF data
format\footnote{Although it was the precursor of the
  \texttt{MathMap} class in AST.}. It was, however, used for two
initially unforeseen purposes which turned out to be very significant:

\begin{enumerate}
\item Associating coordinate systems with plotting surfaces in a
  ``graphics database'' \citep[see \emph{e.g.},][]{SUN48}. This allowed
  plotting applications to store a coordinate system for (say) a graph
  plotted in logarithmic coordinates so that those coordinates could
  later be recovered from the position of a cursor. This demonstrated
  that plotting was a major application area for this type of
  technology, especially when using curvilinear coordinates such as
  Right Ascension and Declination which are notoriously difficult to
  handle properly with standard plotting software.

\item Transformation and combination of bulk image data using general
  arithmetic expressions (as an alternative to combining images using
  a manual sequence of add/subtract/multiply/divide and similar
  applications). This showed that (a) the approach could easily be
  efficient enough to handle large data sets and (b) the data values in
  an image were just another coordinate that could be transformed into
  different representations (logarithmic, different units, \emph{etc}.) in
  much the same way as its axes.
\end{enumerate}

With these insights, it was clear that the ideas behind
\textsc{transform} had potential, but some serious deficiencies had
also emerged:

\begin{itemize}

\item Arithmetic expressions, while fairly general, could not easily
  cope with coordinate transformations that required iterative
  solution, nor with discontinuous transformations, nor with look-up
  tables or a variety of other computational techniques. While
  arithmetic expressions provided a valuable increase in the
  flexibility of coordinate transformations, clearly other classes
  were still needed.

\item It was a major problem for the average writer of astronomical
  software to formulate the required coordinate expressions correctly
  even when dealing with quite simple sky coordinate systems. The core
  of this issue is that celestial coordinate systems are rather
  complex and a good deal of specialist knowledge is needed to
  formulate even simple cases correctly. Clearly a better solution
  would be to encapsulate this knowledge in the WCS software and
  provide a simpler API that dealt only in high-level concepts.

\item For high accuracy work, further complex calculations
  arise. These are related, for example, to atmospheric refraction and
  special \& general relativistic effects (like the observer's motion
  and the sun's gravity).  These require the use of a dedicated
  library of astrometric functions and cannot in practice be handled
  by simple expressions. They also require additional data about the
  observing context (time, position, velocity, wavelength, \emph{etc}.) and
  any practical solution must define how these are stored and
  processed.

\item \textsc{transform} had no ability to store additional
  information about data axes, such as labels and units.

\item It became clear that coordinate transformations frequently
  needed to be combined, for example by applying one transformation
  after another, and that this process was often inefficient. The key
  to better efficiency lay in knowing more about each transformation,
  like whether it was linear or had a variety of other
  properties. With this information it was possible to merge (or
  cancel out) consecutive transformations for better
  efficiency. \textsc{transform} had a rudimentary system for encoding
  this information, but it was not really up to the task.

\item Tying WCS software to a particular (Starlink) data system was a
  mistake and limited the uses to which it could be put. It would
  clearly be better if the data could be encoded (serialised) in
  alternative ways to make it data-system agnostic. The same
  agnosticism should also apply to other likely dependencies, like
  graphics systems and error reporting.  The ability to implement
  these services in alternative ways would be especially important
  when designing graphical user interfaces that processed WCS
  information.

\end{itemize}

\subsection{Developments in FITS WCS}

At about the same time, the wider FITS community also came to recognise
some of the limitations of WCS handling within FITS, and in 1992 work commenced
on a new standard for storing WCS information within FITS files. However,
in view of the ``once FITS, always FITS'' principle \citep[see \emph{e.g.},][]{1997ASPC..125..257W}, that work consisted
mainly in formalising and extending existing practices. So for instance, new
keywords were defined to store the extra meta-data needed for a complete
description of a celestial coordinate system, and new projection types were
added, but the basic model remained unchanged. The new standard still required
that the transformation be split into three components applied in series; an
affine transformation that converts pixel coordinates into \emph{intermediate
world coordinates}, a spherical projection that converts these into \emph{native
spherical coordinates}, and a spherical rotation that converts these into the
final world coordinate system.

In view of the decision to stay with this rigid and restrictive model,
and the expected length of time needed to agree a new standard\footnote{An
expectation that was justified when the standard was finally published in 2002.},
the Starlink project decided in early 1996 to develop its own WCS system, informed
by the earlier experiments with \textsc{transform}, rather than adopt the new FITS
standard.

\subsection{AST Principles}

One of the first decisions was to
separate the representation of a coordinate system (that we called a
Frame) from the computational recipe that transforms between
coordinate systems (a Mapping). From the \textsc{transform} experience
we knew we would need multiple classes of both these data types, all
of which would need to support the same basic operations, but each
having its own specialisation. The correspondence with sub-classing in
object-oriented (OO) programming was irresistible and the decision to
use an OO design immediately followed.

This raised the issue of an implementation language. We planned to use
the SLALIB library for astrometric calculations\footnote{A later version
of AST eventually replaced SLALIB with SOFA and PAL.}. This had been
developed with extreme portability in mind and had recently been
re-written in ANSI C. We didn't want to compromise this portability,
so decided also to work in strict ANSI C and to minimise software
dependencies as much as possible. This meant providing portable
interfaces to facilities that were intrinsically less portable, such
as data file access, plotting, error reporting, \emph{etc}. and providing
simple implementations that users could re-write if necessary.

Deciding to write an OO system in a non-OO language took considerable
thought. We needed to provide a Fortran-callable interface but, at the
time, the portability of C++ code was quite limited if one needed to
call it from Fortran, so that route was unattractive\footnote{Further, there
was no official C++ standard available at the time - the first international
standard for C++ (ISO/IEC 14882:1998) was published in 1998.}. Eventually, we
were guided by the approach described by \citet{1992Holub} for
handling objects in C and were able to hide the detail from users
using pre-processor macros.

One consequence of this is that users cannot easily create new
sub-classes from AST objects without learning the internal conventions
that it uses. At the time, this was seen as something of an advantage.
The library is intended for data interchange and creating new
sub-classes would inevitably allow persistent objects to be created
that other users could not access. However, with hind-sight a more open
architecture may have encouraged involvement from a wider user-base
\footnote{For Mappings, where this
  issue is most relevant, the problem has been mitigated in a
  controlled way by the IntraMap class that allows separately-compiled
  code to be imported into the library.}.

As noted previously, we wanted the AST API to deal in high-level
concepts and to hide as much specialist detail as possible from the
user. This principle arose from considering the complex calculations
involved in handling celestial coordinates and time systems. However,
we soon realised that two other areas were similarly complex and could
benefit from the same approach.

The first area was graphics. Plotting in curvilinear coordinates is a
complicated business if one wants to handle all the corner cases
correctly. Plotting and labelling celestial coordinate axes, for
example, presents many problems; especially near the poles of an all-sky
projection. It is made even harder if the projection contains
discontinuities. But the high-level concepts involved in such plotting
(coordinate systems and the mappings between them) are such a natural
fit with other AST concepts that it seemed obvious to implement a class
of coordinate system that is specialised for graphics. The high-level
operations it supports would then hide the details of the complex and
generalised plotting algorithms involved.

The second area is an aspect of data storage -- namely the handling of
FITS header cards. While the AST library could provide ways to
serialise its own data transparently, possibly in multiple ways, it
also needed to inter-operate with FITS. WCS data in FITS data files is
stored in a series of 80-character header cards and, over the years,
the number of different ways the information can be stored in these
headers has grown. The complexity involved is now considerable
\citep[see \emph{e.g.},][]{2015Thomas}. Again, detailed specialist knowledge
is needed to extract this information reliably and to write it back
(possibly modified) in a form that gives other FITS-handling software
a chance of using it while not conflicting with the many other FITS
headers typically present.

For a user of the AST library, we wanted this process of accessing FITS
headers to appear as much like a simple read/write operation as
possible, with all the implementation details hidden. This requirement
arose from more than simply ease-of-use. FITS header conventions (many
of them informal) are in constant flux and if these details are embedded
in applications programs, those programs must constantly receive
attention if they are to remain up to date. Embedding all these details
in the AST library allows the problem to be addressed in one place and
by someone with the necessary expertise.

FITS header handling has proved one of the most complex areas to tame in
AST. But introducing the concept of a \emph{destructive read} (which
reads WCS data from FITS headers and simultaneously deletes the relevant
headers) has made it possible to write applications with very little code
that have completely general handling of FITS WCS headers (see
section~\ref{sec:fitsencodings}).

\section{The AST Data Model}
\label{sec:model}

The most basic principle behind the AST data model is a clear distinction
between a \emph{transformation} and a \emph{coordinate system}.

A \emph{transformation} is a mathematical recipe for converting a numerical
input vector into a corresponding numerical output vector. The
transformation itself has no knowledge of what the values within these
vectors represent, other than that each one constitutes a position within
some unspecified N-dimensional space\footnote{The dimensionality of the
output space need not be the same as that of the input space.}.

A \emph{coordinate system} is a collection of meta-data describing a set of
one or more axes. This will include:
\begin{itemize}
\item the number of axes (\emph{i.e.}, dimensionality of the space)
\item the physical quantity described by each axis
\item the units used by each axis
\item the geometry of the space that they describe (flat, spherical,
\emph{etc})
\item the nature of the coordinate system (Cartesian, polar, \emph{etc})
\item other meta-data that may be needed to specify the coordinate system
fully.
\end{itemize}

These two concepts are encapsulated within two separate classes in the
AST data model: the \emph{Mapping} class and the \emph{Frame} class. This
separation underlines the fundamental difference between the two main
requirements of any coordinate handling system:

\begin{enumerate}
\item knowing \emph{how} to convert numerical positions from one coordinate
system to another.
\item knowing \emph{what} those coordinate systems represent.
\end{enumerate}

As an example of the practical consequences of this distinction,
the \emph{pixel size} of a typical 2-dimensional image of the sky is
\emph{not} considered to be a property of the (RA,Dec) Frame, since it is
determined by the nature of the transformation from pixel coordinates to
(RA,Dec) coordinates, rather than being an intrinsic property of the (RA,Dec)
coordinate system itself.

The two classes, \emph{Mapping} and \emph{Frame}, are extended
to create a wide variety of sub-classes, each of which describes a
specific form of transformation or coordinate system. New sub-classes can
be added as required and slot naturally into the existing
infra-structure provided by the rest of AST.

In addition, this separation into two ``orthogonal'' classes makes it
easy to create complex compound objects from simple component objects.
For instance, multiple Mappings can be combined into a new object, and
the resulting object will itself be a Mapping. Likewise, multiple Frames
can be combined into a new object, and the resulting object will itself be
a Frame\footnote{For instance, a 2-dimensional (RA,Dec) Frame can be
combined with a 1-dimensional wavelength Frame to create a 3-dimensional
(RA,Dec,Wavelength) Frame.}.

\begin{figure}[h]
\centering
\includegraphics[width=0.5\columnwidth]{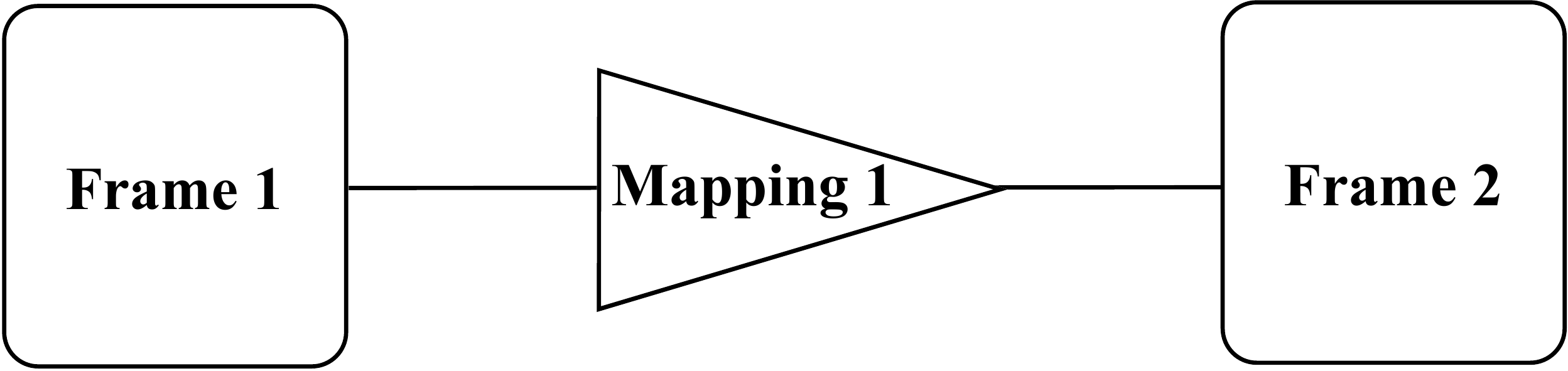}
\caption{A FrameSet that describes two coordinate systems and the
transformations between them. The Mapping's \emph{forward} transformation
transforms positions in \emph{Frame 1} to the corresponding position in
\emph{Frame 2}. The Mapping's \emph{inverse} transformation
transforms positions in \emph{Frame 2} to the corresponding position in
\emph{Frame 1}. }
\label{fig:simple-frameset}
\end{figure}

However, at some point these two classes need to be brought together to
provide a complete description of a set of related coordinate systems.
The \emph{FrameSet} class is used for this purpose. A \emph{FrameSet}
encapsulates a collection of two or more Frames, with the Mappings that
describe the transformation between the corresponding coordinate systems.
The simplest FrameSet contains two Frames, together with a single Mapping that
describes how to convert positions between these two Frames (see
Fig.~\ref{fig:simple-frameset}).

\begin{figure}[h]
\centering
\includegraphics[width=\columnwidth]{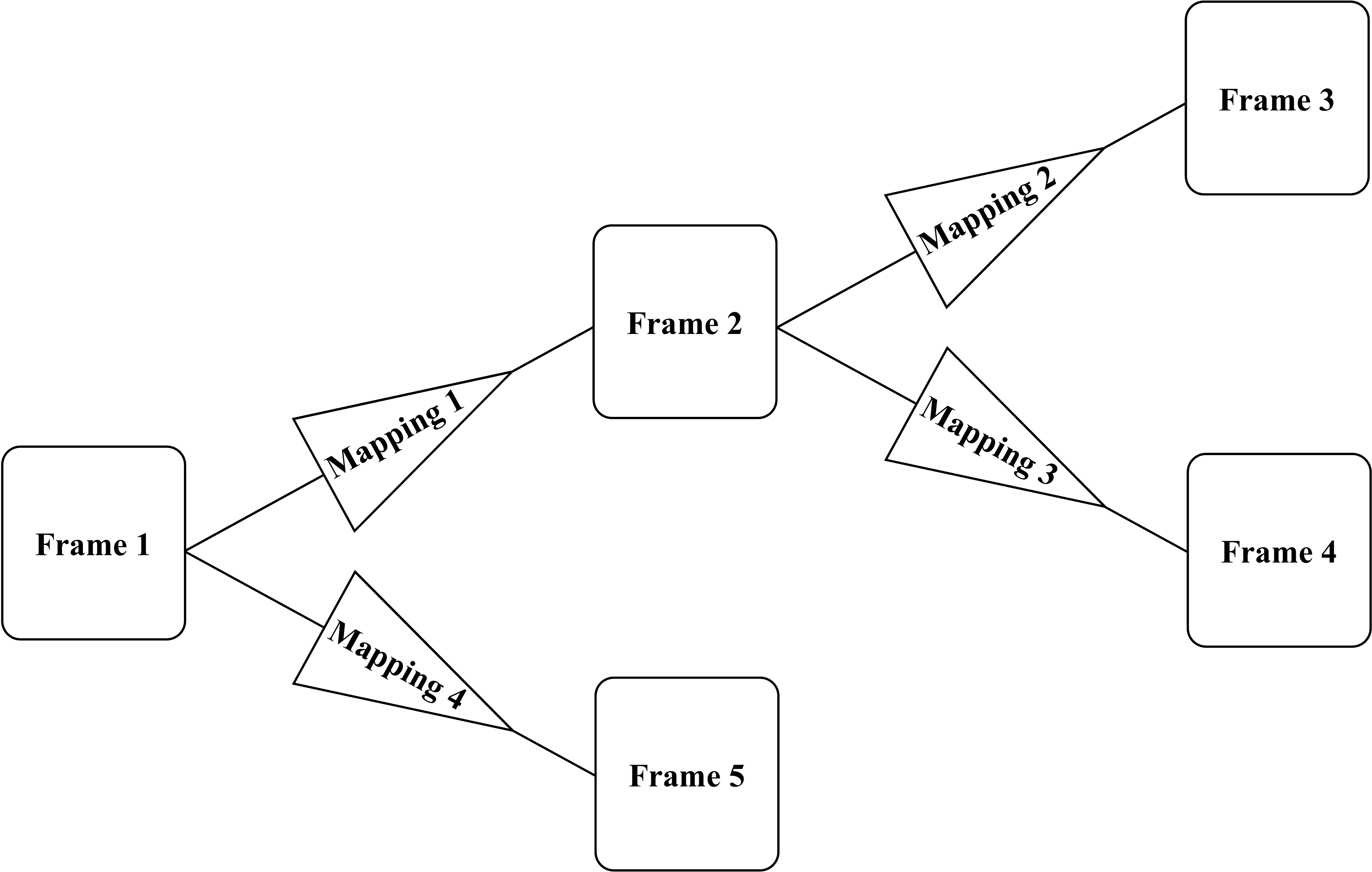}
\caption{A FrameSet that describes five inter-related coordinate systems
and the transformations between them. }
\label{fig:complex-frameset}
\end{figure}

More complex FrameSets can be created that describe the relationships between
multiple Frames in the form of a tree structure (see
Fig.~\ref{fig:complex-frameset}).

\subsection{Transformations and Mappings}
\label{sec:mappings}
Within AST, most \emph{Mappings} encapsulate two transformations - one is
designated as the \emph{forward} transformation and the other as the
\emph{inverse} transformation. When a Mapping is used to transform a set
of positions, the caller must indicate if the forward or inverse
transformation is to be used. The \emph{forward transformation} converts
positions within the input space of the Mapping into corresponding
positions within the output space, and the \emph{inverse transformation}
converts positions within the output space of the Mapping into corresponding
positions within the input space. A Mapping can be \emph{inverted}, which
results in the two transformations being swapped.

For most classes of Mapping, the inverse transformation is a genuine
mathematical inverse of the forward transformation. However, this is not
an absolute requirement, and there are a few classes of Mapping where
this is not the case (for instance the \emph{PermMap} class, when the
axes of the output space are a permuted subset of the axes of the input
space). In addition, the Mapping class does not require that \emph{both}
transformations are defined. For instance, the \emph{MatrixMap} class, which
multiplies each input vector by a specified matrix to create the output
vector, will only have an inverse transformation if the matrix is square
and invertable.

\subsubsection{Atomic Mappings Provided by AST}
AST provides many classes of Mapping that implement a wide range of
different transformations. Most of these are \emph{atomic} Mappings that
implement a specific numerical transformation and, if possible, its inverse.
But some are \emph{compound} Mappings that combine together other Mappings
(atomic or compound) in various ways to create a more complex Mapping.
A compound Mappings does not define its own transformations, but instead
inherits the transformations of the individual component Mappings which
it encapsulates.

The most significant atomic Mapping classes are:

\begin{description}
\item[UnitMap:] Copy positions from input to output without any change.
\item[WinMap:] Transform positions by scaling and shifting each axis.
\item[ZoomMap:] Transform positions by zooming all axes about the origin.
\item[ShiftMap:] Translate positions by adding an offset to each axis.
\item[MatrixMap:] Transform position vectors by multiplying by a matrix.
\item[PolyMap:] A general N-dimensional polynomial transformation.
\item[PermMap:] Transform positions by permuting and selecting axes.
\item[LutMap:] Transform 1-dimensional coordinates using a look-up table.
\item[MathMap:] Transform coordinates using general algebraic mathematical expressions.
\item[WcsMap:] Implements a wide range of spherical projections.
\item[SlaMap:] Transform positions between various celestial coordinate systems.
\item[SpecMap:] Transforms positions between various spectral coordinate systems.
\item[SphMap:] Map 3-d Cartesian to 2-d spherical coordinates.
\item[TimeMap:] Transform positions between various time coordinate systems.
\item[PcdMap:] Apply 2-dimensional pincushion/barrel distortion.
\item[DssMap:] Transform positions using Digitised Sky Survey plate solutions.
\item[GrismMap:] Models the spectral dispersion produced by a grism.
\item[IntraMap:] Transform positions using an externally supplied transformation function.
\item[SelectorMap:] Locates positions within a set of Regions (see
section~\ref{sec:cmpmap}).

\end{description}

All classes of Mapping are immutable. That is, a Mapping cannot be changed
once it has been created. This is unlike Frame and FrameSet objects, which
\emph{can} be changed.

\subsubsection{Compound Mappings}
\label{sec:cmpmap}

The two most significant compound Mapping classes are:

\begin{description}
\item[CmpMap:]  The CmpMap class is a subclass of Mapping that encapsulates
two other Mappings either in series or in parallel. Either or both of the two
encapsulated Mappings can itself be a CmpMap, allowing arbitrarily complex
Mappings to be created.

In a \emph{series} CmpMap, each input position is transformed by the
first component Mapping, and the output from that Mapping is then
transformed by the second component Mapping. Consequently, the output
dimensionality of the first Mapping must be the same as the input
dimensionality of the second Mapping.

In a \emph{parallel} CmpMap, the input space is split into two sub-spaces.
The first component Mapping is used to transform the axis values
corresponding to the first subspace, and the second component Mapping is
used to transform the axis values corresponding to the second subspace.
Thus the input dimensionality of the CmpMap is equal to the sum of the
input dimensionalities of the two component Mappings, and the output
dimensionality of the CmpMap is equal to the sum of the output
dimensionalities of the two component Mappings.

\item[SwitchMap:] The SwitchMap class is a subclass of Mapping that allows
a different transformation to be used for different regions within the input
space.

Each SwitchMap encapsulates any number of other Mappings, known as ``route''
Mappings, and one ``selector'' Mapping. All of these Mappings must have the
same input dimensionality, and all the route Mappings must have the same
output dimensionality. The selector Mapping must have a one-dimensional
output space.

Each input position supplied to the SwitchMap is first transformed by the
selector Mapping. The scalar output from the selector Mapping is used to
index into the list of route mappings. The selected route mapping is then
used to transform the input position to generate the output position
returned by the SwitchMap.

There is a specialised subclass of Mapping, the SelectorMap class, that
is designed specifically to fulfil the role of the selector Mapping
within a SwitchMap, but in principle \emph{any} suitable form of Mapping
may be used. The SelectorMap class encapsulates several Regions (see
section~\ref{sec:region}) and returns an output value that indicates
which of the  Regions (if any) contained the input value. Thus, the
SwitchMap would typically contain one route Mapping for each of the
Regions contained within the SelectorMap.

\end{description}

\subsubsection{Simplification}
\label{sec:simplification}

There are a wide range of possible transformations that could
potentially be applied to a data set during analysis. These
include simple things such as rotation, scaling, shear, \emph{etc}., but
could in principle include more complex transformations such as
re-projection, dis-continuous ``patchwork'' transformations, or even
transformation using a general algebraic expression.  A coordinate
handling system should make it possible for a user to apply an
arbitrary set of such transformations in series to a data set, without
losing track of the coordinates of each data point. With a
prescriptive scheme such as FITS-WCS this would require each
transformation to locate the appropriate component of the FITS-WCS
pixel to world coordinate mapping, and modify the corresponding
headers in a suitable way. This is often a difficult, if not
impossible, task. Within AST, the chaining of transformations is
accomplished simply by creating a Mapping that describes each new
transformation and concatenating it with the existing pixel to world
coordinate mapping by creating a new CmpMap (see section~\ref{sec:cmpmap}).

However, by itself this can lead to Mappings that become increasingly
complex as transformations are stacked on top of each other. This is
a problem because it leads to:

\begin{enumerate}
\item slower evaluation of the total transformation,
\item less accurate evaluation of the total transformation, and
\item more room being needed for storage.
\end{enumerate}

To avoid this, the Mapping class provides a method that takes a
potentially complex Mapping and simplifies it as far as possible. Doing
such simplification in a general and effective manner is one of the most
difficult challenges faced by the AST model, but experience has shown
that the current scheme handles most cases
sufficiently well. The steps involved in simplification depend on the
nature of the component Mappings in the total CmpMap. Each class of
Mapping provides its own rules that indicate when and how it can be
simplified, or combined with an adjacent Mapping in the chain. To
illustrate the principle, some of the simplest examples include:

\begin{enumerate}
\item any Mapping can be combined with its own inverse to create a UnitMap,
\item UnitMaps can be removed entirely,
\item adjacent MatrixMaps in series can be combined using matrix
multiplication to create a single MatrixMap,
\item adjacent MatrixMaps in parallel can be combined to create a
single MatrixMap of higher dimensionality (filling the off-diagonal
quadrants with zeros),
\item adjacent ShiftMaps can be combined to form a single ShiftMap
(either in series or in parallel).
\end{enumerate}

The whole simplification process is managed by the CmpMap class - it
expands the compound Mapping into a list of atomic Mappings to be applied
in series or parallel, and then for each Mapping in the list, invokes
that Mapping's protected \texttt{astMapMerge} method. This method is
supplied with the entire list of atomic Mappings, and determines if the
nominated Mapping can be merged with any of its neighbours. If so, a new
list of Mappings is returned containing the merged Mapping in place of
the original mappings. Once all atomic Mappings in the CmpMap have been
checked in this way, the same process is repeated again from the
beginning in case any of the changes that have been made to the list
allow further simplifications to be performed. This process is repeated
until no further simplifications occur.

There are in general multiple ways in which a list of Mappings (either
series or parallel) can be simplified. Since each class of Mapping has
its own priorities about how to merge itself with its neighbours, it is
possible for the simplification process to enter an infinite loop in
which neighbouring Mappings disagree about the best way to simplify the
Mapping list. When a particular Mapping is asked to merge with its
neighbours, it may make changes to the Mapping list that are then undone
when a neighbouring Mapping is asked to merge with its neighbours. The
simplification method takes care to spot such loops and to assign priority
to one or the other of the conflicting Mappings.

\subsubsection{Missing or Bad Axis Values}
\label{sec:bad}
AST flags unknown or missing axis values using a special numerical value,
known as the ``bad'' value. If an input position supplied to a Mapping
contains one or more bad axis values, then in general all output axis
values will be bad\footnote{One exception is that a parallel CmpMap may
be able to generate non-bad values for some of its output axes.}.

Mappings may also generate bad output axis values if the input position
corresponds to a singularity in the transformation, or is outside the
region in which the transformation is defined.

\subsection{Frames and Domains}
As described in section~\ref{sec:model}, each instance of the Frame class
contains all the meta-data necessary to give a complete description of
a particular coordinate system. Each Frame is associated with a specific
\emph{domain} as explained in the next section.

Unlike Mappings, the properties of a Frame can be changed at any time.

\subsubsection{What is a Domain?}
\label{sec:domain}
AST uses the word \emph{domain} to refer to a physical (or abstract)
space such as ``time'', ``the sky'', ``the electro-magnetic spectrum'',
``the focal plane'', ``a pixel array''. Points within such a space can in
general be described using any one of several coordinate systems. For
instance, any position on the sky can be described using ICRS
coordinates, Galactic coordinates, \emph{etc}. Similarly, positions in
the electro-magnetic spectrum can be described using frequency,
wavelength, velocity, \emph{etc}.

Each subclass of Frame represents a specific domain with a domain name (a
string) which is the same for all instances of the class. However, each
instance of the class represents just one particular coordinate system
within that domain. In general, each Frame sub-class will also
encapsulate all the information needed to create a Mapping between any
pair of supported coordinate systems within its domain.

For example, the \emph{SkyFrame} class represents the astronomical sky (its
domain) and has the domain name ``SKY''. It knows about a range of celestial
coordinate systems that can be used to describe positions on the sky.
However, an instance of the SkyFrame class represents only one of these
celestial coordinate systems at a time. The particular one it represents is
stored in its ``System'' attribute (an instance variable) which takes
values such as ``ICRS'', ``FK5'', ``Galactic'', \emph{etc.}

The SkyFrame class extends the Frame class by incorporating various other
items of metadata necessary to perform conversion between the supported
celestial coordinate systems, the main items being the epoch of observation
and the reference equinox \footnote{The base Frame class includes the
observer's geodetic position.}. These are also instance variables and
therefore specific to each SkyFrame instance. The coordinate system
represented by an instance of the SkyFrame class can be changed at any time
simply by assigning new values to any of the relevant attributes, as in
the following Python example:

\begin{lstlisting}
import starlink.Ast as Ast

# Create a Frame describing ICRS Right
# Ascension and Declination. This is the
# default system for the SkyFrame class.
frame1 = Ast.SkyFrame()

# Create a deep copy, and change the System
# attribute of the copy so that it represents
# Galactic longitude and latitude.
frame2 = frame1.copy()
frame2.System = "Galactic"
\end{lstlisting}

\subsubsection{Coordinate Conversion within Domains}
\label{sec:domConversion}
Given the two SkyFrames from the example in the previous section, it is
possible to create a Mapping between them (\emph{i.e.}, from ICRS (RA,Dec)
to Galactic ($\ell,b$)) based on the meta-data stored within the two
SkyFrames. This Mapping can then be used to convert (RA,Dec) positions
into equivalent (l,b) positions, or vice-versa. The process of creating
this Mapping is implemented within the \texttt{astConvert} method of the
Frame class. For instance:

\begin{lstlisting}
my_frameset = frame1.convert(frame2)
\end{lstlisting}

will, if possible, generate a Mapping from \texttt{frame1} to
\texttt{frame2}. The returned Mapping is encapsulated within a FrameSet that
also includes copies of \texttt{frame1} and \texttt{frame2}.

Likewise, the SpecFrame class encapsulates all the information needed to
create Mappings between any pair of supported spectral coordinate system,
accounting for rest frequency, standard of rest, celestial reference position,
\emph{etc.}, as in the following example:

\begin{lstlisting}

#  Create a Frame describing helio-centric
#  Radio Velocity in units of km/s, with a rest
#  frequency of 345.8 GHz.
frame1 = Ast.SpecFrame("System=VRAD,RestFreq=345.8")

#  Create a deep copy, and change the attributes
#  so that the copy describes frequency units of
#  "Hz" with respect to the kinematic Local Standard
#  of Rest.
frame2 = frame1.copy()
frame2.System = "FREQ"
frame2.Unit_1 = "Hz"
frame2.StdOfRest = "LSR"

#  Create a FrameSet that contains the Mapping
#  between these two spectral coordinate systems.
my_frameset = frame1.convert(frame2)
\end{lstlisting}

The principle that each class of Frame contains all the meta-data and
intelligence required to create a Mapping between any two coordinate
systems within the Frame's domain, extends to compound Frames as well as
atomic Frames, as described in section~\ref{sec:comConversion}.

The base Frame class itself is slightly unusual in that it can be used to
describe any generic domain, and is restricted to a single Cartesian
coordinate system within that domain. The domain associated with a basic
Frame is specified by the caller and can be any arbitrary string. Clearly,
this restricts the usefulness of a basic Frame (compared to more
specialised classes of Frame) in that it is not possible to include any
knowledge about multiple coordinate systems given the arbitrary nature of
the domain. The only exception is that the basic Frame class
knows how to convert between different dimensionally equivalent units.
Thus the implementation of the \texttt{astConvert} method provided by the
basic Frame class can generate a Mapping between two basic Frames if they
have the same domain name, the same number of axes, and the axes have
dimensionally equivalent units.

\subsubsection{Atomic Frame Classes Provided by AST}
AST provides several classes of Frame that describe different specialised
domains. As with Mappings, these can be divided into \emph{atomic} Frames that
describe a single specific domain, and \emph{compound} Frames that combine
together other Frames (atomic or compound) to create a Frame describing a
domain of higher dimensionality. The atomic Frame classes are:

\begin{description}
\item[Frame:] An arbitrary N-dimensional domain with a single Cartesian
coordinate system.
\item[FluxFrame:] A 1-dimensional domain describing several forms of flux
measurement systems (all measured at a single spectral position). In this
case, the ``axis value'' represents a flux value.
\item[SkyFrame:] A 2-dimensional domain describing several celestial
coordinate systems.
\item[SpecFrame:] A 1-dimensional domain describing several spectral
coordinate systems.
\item[DSBSpecFrame:] Extends the \emph{SpecFrame} class to describe dual
sideband spectral coordinate systems.
\item[TimeFrame:] A 1-dimensional domain describing several time coordinate
systems.
\end{description}

\subsubsection{Compound Frames}

The following compound Frame classes are provided:

\begin{description}

\item[CmpFrame:]  The \emph{CmpFrame} (compound Frame) class describes a
coordinate system that combines the axes from two other Frames, in any
order. The name of the domain associated with a CmpFrame is constructed
automatically from the domain names of the two component Frames. For
instance if a CmpFrame contains a SkyFrame and a SpecFrame, then its
domain name will be ``SKY-SPECTRUM''.

\item[SpecFluxFrame :] The \emph{SpecFluxFrame} class combines a FluxFrame
with a SpecFrame.

\end{description}

\subsubsection{Coordinate Conversion within Compound Domains}
\label{sec:comConversion}
The principle that Frame classes contain all the information necessary to
convert between any coordinate systems within their domain also applies
to compound Frames and the compound domains that they represent. This
facility is accessed through the \texttt{astConvert} method, as with atomic Frames
(see section~\ref{sec:domConversion}).

With atomic Frames, conversion is only possible between instances of the
same Frame class (\emph{i.e.}, Frames that represent alternative coordinate
systems within the same physical domain). However, with a compound Frame
containing (say) two sub-components, it may be possible to find a
conversion to either or both of the two sub-Frames, so that a new range
of possibilities opens up. For instance, a CmpFrame describing the
3-dimensional ``SKY-SPECTRUM'' domain might be matched with any of the
following:

\begin{itemize}
\item Another CmpFrame describing the SKY-SPECTRUM domain.
\item Another CmpFrame describing the SPECTRUM-SKY domain.
\item A SkyFrame (\emph{i.e.}, a Frame describing the SKY domain).
\item A SpecFrame (\emph{i.e.}, a Frame describing the SPECTRUM domain).
\end{itemize}

where a ``match'' means that conversion is possible. If the destination
Frame has fewer axes than the source Frame, then the resulting Mapping
will contain a PermMap - an atomic Mapping that permutes and selects a
subset of its input axes.

Because CmpFrames can be nested arbitrarily deeply and their axes can be
permuted, deciding whether conversion is possible can be non-trivial, but
the facility has great power and can form the basis of a search function
for coordinate systems. For instance, if a general purpose program reads
the WCS from a data file of arbitrary dimensionality and wants to ask the
question ``can I determine the ICRS (RA,Dec) of each pixel position?'' it
can create a SkyFrame describing ICRS (RA,Dec) and then use the
\texttt{astConvert} method to see if a Mapping can be created from the
WCS Frame read from the data file to the \emph{template} SkyFrame
describing the required coordinate system. This will allow ICRS (RA,Dec)
to be determined for any 2-dimensional data file calibrated in any of the
supported celestial coordinate systems, and also for any multi-dimensional data
file that contains a pair of celestial coordinate axes.

\subsubsection{Other Important Frame Methods}
A primary goal of AST is to support high-level software that can process
arbitrary coordinate systems without needing to know their details. This
simplifies the high-level algorithms themselves and also allows them to
benefit from later additions to the AST library (\emph{e.g.}, new coordinate
systems and domains) that did not exist when they were originally
written. The object-oriented design of AST is key to this feature and it
particularly affects the Frame classes.

To this end, the base Frame class declares a range of methods that
provide a generic interface for handling coordinate systems. These
methods are then typically over-ridden by sub-classes so that they
exhibit the required more specialised behaviour when manipulated by
generic high-level algorithms. Rather than detail the interface in full,
we will illustrate the principle with a few examples here.

Encoding and decoding of axis values to/from text is a function that
differs in detail between Frame sub-classes because of the differing
conventions for formatting, say, simple floating point values, angles and
times (the last two having a wide range of possible representations). The
base Frame class defines a decoding interface that operates on arbitrary
input text and converts it into axis values, regardless of the nature of
the formatting involved or, indeed, which characters and delimiters are
present. With suitable over-rides to the implementation in Frame
sub-classes, this interface allows high-level software to read
free-format input containing axis data for arbitrary Frames, including
compound Frames, using a simple algorithm.

A matching encoding interface formats axis values under the control of a
``Format'' attribute string whose syntax can be over-ridden in Frame
sub-classes to suite the formatting required. For example, the C-like
``\texttt{\%.4f}'' might describe the formatting for a simple floating
point axis, while the more specialised ``\texttt{hh:mm:ss}'' could be
used for a Right Ascension axis. Being Frame-specific, however, these
strings need to be set by specialised software, although defaults are
provided. For a more generic interface, a ``Digits'' attribute is
available that controls the number of significant digits shown and
applies to any class of Frame. This finds particular use during graphical
operations (see section~\ref{sec:plotting}).

A related method defined by the base Frame class will wrap axis values
into a standard range (\emph{e.g.}, 0 to 360 degrees) where appropriate.
This can be useful when reading and writing axis values, but also
provides a simple way to characterise the different topologies of (say)
flat, cylindrical and spherical spaces. A further method will return the
distance between pairs of points, thereby defining the metric of the space
that a Frame represents. The inverse operation is to return the
coordinates of a point which is offset a given distance from a starting
position in a defined direction.

These methods, suitably over-ridden in each Frame sub-class, form the
basis for many generic graphics operations such as drawing geodesic
curves (which replace straight lines as the basic drawing element in AST)
and coordinate axes. For example, to mark formatted numerical values on
the possibly curved axis of a graph, the encoding function described
above is used. The formatting precision (``Digits'' attribute) is
progressively increased until all the formatted label values become
distinct. This ensures adequate, but not excessive precision regardless
of the range of axis values present and the algorithm works regardless of
the type of Frame involved.

\subsection{FrameSets}

Converting between different coordinate systems within the same domain is
handled by Frame and its sub-classes. However, converting between
different domains (\emph{e.g.}, between pixel coordinates and sky coordinates) is
not possible in this way because there is no intrinsic relationship
between the two domains unless extra information is supplied. This extra
information typically describes experimental configurations; for example
where a telescope was pointing (or the settings of a spectrograph). Given
this additional detail, it becomes possible to tie the two domains
together, so that conversion can be performed between a pixel-based
coordinate system and any celestial coordinate system (or wavelength
system in the case of a spectrum).

Linking domains together in this way essentially creates a ``super
domain'' where otherwise unconnected coordinate systems are bound
together in a similar relationship to coordinate systems within a single
domain.  This facility is typically required when calibrating the
coordinate systems of experimental data and is provided by the
\emph{FrameSet} class. It also finds use in graphics applications for
attaching a variety of (typically curvilinear) coordinate systems to a
plotting surface.

Each domain within this ``super domain'', is represented by a separate
Frame. If a FrameSet is used, for instance, to provide a complete
description of the WCS associated with a data array, then one of the
Frames within the FrameSet will represent pixel coordinates, and the
other Frames will represent a collection of alternative world coordinate
systems. For instance, a FrameSet describing an image taken by a
telescope may have three Frames describing pixel coordinates, focal plane
coordinates and sky coordinates.

\subsubsection{FrameSets as Tree Structures}
A FrameSet represents a network of inter-related coordinate systems in
the form of a tree-structure in which each node is a Frame, with the
nodes being connected together by Mappings (see Fig.~\ref{fig:complex-frameset}).

The FrameSet class provides a method that returns the Mapping between any
two nominated Frames. This may involve concatenating several Mappings if
the two Frames are not directly connected to each other. For instance, if
asked to return the Mapping between Frame 1 and Frame 4 in
Fig.~\ref{fig:complex-frameset}, the method will retrieve Mappings 1 and
3 from the FrameSet, and combine them in series into a single
\emph{CmpMap} (compound Mapping).

The FrameSet class also provides a method that returns any nominated Frame
from a FrameSet.

New Frames can be added to an existing FrameSet at any point in the tree
structure. To do so, the caller must provide a Mapping that maps positions
from an existing Frame in the FrameSet to the new Frame.

Frames can also be removed from a FrameSet. If the Frame is a leaf node
in the tree structure, then it is simply removed, together with the
Mapping that  connects it to its parent Frame. If the Frame is \emph{not}
a leaf node, the Frame is removed but the Mapping that connects it to its
parent Frame is retained so that its child Frames can still be reached.

\subsubsection{Base and Current Frames}
\label{sec:basecurrent}
Two of the most common operations on FrameSets are 1)
converting positions from one Frame to another, and 2) enquiring or using
the properties of a Frame. Performing these two operations repeatedly
would become tedious if each such operation involved separate calls
to extract the required Mapping or
Frame from the FrameSet. To avoid this, AST is implemented in such a way
that the FrameSet class effectively inherits from both the Mapping class
and the Frame class. This means that any Mapping method, or any Frame
method, can also be used on a FrameSet.

The FrameSet class allows two nominated Frames to be flagged within each
FrameSet. One is referred to as the ``current Frame'', and the other as
the ``base Frame''\footnote{The FrameSet class provides methods to set
and retrieve the index of these two Frames.}. When used as a Frame, a
FrameSet is equivalent to its current Frame. When used as a Mapping, a
FrameSet is equivalent to the Mapping from its base Frame to its current
Frame.

When attaching WCS information to other data a FrameSet is typically
employed and, by convention, its base Frame is used to represent the
intrinsic coordinate system associated with the data. For example, the
base Frame may be identified with the pixel coordinate system of an image
or with the native coordinates used to address a plotting surface. The
other Frames in the FrameSet represent alternative coordinate systems
with which a user may choose to work and the current Frame represents the
currently-selected coordinate system.

For example, an application that locates objects in images may generate
results in pixel coordinates, but can then transform them between the
base and current Frame coordinate systems in order to present them to the
user in the required form (possibly as celestial coordinates). Similarly,
a cursor position read from a graphics device can be shown to the user
not in device coordinates, but in a form they can more readily
understand.

When AST reads WCS data from a FITS file, it creates a FrameSet with this
form; the base Frame represents pixel coordinates and the current Frame
represents the FITS primary world coordinate system. The FrameSet may
also contain other Frames representing any alternate axis descriptions
stored in the FITS-WCS. This means that the FrameSet can be used as a
Mapping from pixel coordinates to primary WCS, and can also be used as a
Frame to determine the properties of the primary WCS.

The caller is free to select new base and/or current Frames at any time.

\subsubsection{Integrity Restoration}
\label{sec:integrity}
Consider the simple case mentioned above where a FrameSet is used to
describe the WCS in a 2-dimensional image. The FrameSet could for
instance contain a pixel Frame as the base Frame and an (RA,Dec) Frame as
the current Frame, connected together by a suitable Mapping. It is
clearly possible to break the integrity of such a FrameSet, such that the
Mapping no longer accurately describes the transformation from pixel
coordinates to world coordinates. One obvious way in which this could be
done is to change the current Frame so that it describes, say, Galactic
coordinates rather than (RA,Dec). The Mapping is left unchanged and so will
still generate (RA,Dec) values, even though the FrameSet now claims that
these are galactic coordinates\footnote{Note, Mappings are immutable and so the
integrity of a FrameSet cannot be broken by making changes to a Mapping.}.

In order to retain the integrity of the FrameSet, the Mapping must
be replaced with one that generates the appropriate galactic coordinate
values rather than (RA,Dec) values. One way in which this could be done is as
follows, starting from the original unmodified FrameSet:

\begin{enumerate}
\item Create a copy of the current Frame (\emph{i.e.}, the (RA,Dec) Frame), and
change the attributes of the copy so that it describes Galactic
coordinates.
\item Use the \texttt{astConvert} method on the (RA,Dec) Frame to generate a
Mapping from (RA,Dec) to galactic coordinates.
\item Add the galactic coordinates Frame into the FrameSet, using the
above Mapping to connect it to the existing (RA,Dec) Frame.
\item Remove the original (RA,Dec) Frame.
\end{enumerate}

The final FrameSet is unchanged in the sense that it still contains two
Frames, but now the Mapping that connects them correctly generates the values
described by the new current Frame (\emph{i.e.}, Galactic coordinates).

However, the above process is quite involved and prone to error, and so
the FrameSet class itself provides generalised ``integrity restoration''
along the same lines, meaning that client code is relieved of the
responsibility.

In summary, the integrity restoration system within the FrameSet class
means that whenever the properties of the current Frame are changed via a
FrameSet reference, the Mappings within the FrameSet are automatically
modified accordingly. However, this mechanism may be circumvented when
necessary by first obtaining a direct reference to an internal Frame
within the FrameSet and making the changes via this reference. In this
case, the Mappings within the FrameSet are left unchanged.

\subsubsection{Searching a FrameSet for a Frame with Required Properties}
As described earlier, the \texttt{astConvert} method defined by the Frame
class attempts to find a Mapping between two arbitrary
Frames\footnote{This may or may not be possible depending on the nature
of the two Frames.}. Since the FrameSet class inherits the methods of the
Frame class, the \texttt{astConvert} method can also be used on FrameSets. In
this case the \texttt{astConvert} method will search through all the Frames
in the FrameSet, starting with the current Frame, until a Frame is found
for which a Mapping can be created. Since, in general, the Frames within a
FrameSet all describe different domains, it is unlikely that more than one
Frame will generate a Mapping, but the search order can be controlled by
the caller in order to assign priority to specific Frames.

This allows code to search a FrameSet for a Frame that has specific
properties. For instance, a Frame can be created with the required
properties and then used as a ``template'' to search a FrameSet:

\begin{lstlisting}
template = Ast.SkyFrame()
result = frameset.convert(template)
if result is None:
  print("No celestial coord system found")
\end{lstlisting}

In this example \texttt{frameset} is searched for a SkyFrame. If found, a
new FrameSet is returned in which the base Frame is the matching Frame
from \texttt{frameset}, the current Frame is a copy of \texttt{template} and
the appropriate Mapping exists between them. In addition, the base Frame
of \texttt{frameset} is set to indicate the matching Frame. If no matching
Frame is found, a null reference is returned by \emph{astConvert}.

In the above example the specific celestial coordinate system represented
by \texttt{template} was left unspecified when the SkyFrame was created.
Consequently, the copy of \texttt{template} included in the
\texttt{result} FrameSet returned by \emph{Convert} inherits the
celestial coordinate system of the matching Frame within
\texttt{frameset}. If a specific system was specified when \texttt{template}
was created, then that system would be given priority and be included in
the returned \texttt{result} FrameSet.

\subsection{Regions}
\label{sec:region}
Usually, axis values are stored simply as floating point values or arrays
in the user's programming language and are not bound into AST objects.
This is because the values are normally obtained and later used in this
form (in other parts of the software, independently of AST). Also, speed
of processing is vital. If axis values were routinely wrapped inside AST
objects, the overhead of wrapping and unwrapping would be considerable,
especially for small sets of positions that are processed repeatedly.

Nevertheless, axis values strictly only make sense within a particular
coordinate system and this association can be made explicit, if required,
by binding a set of positions to a specific Frame. This is performed by
the Region class which encapsulates a Frame and a list of points,
specified by their positions in the coordinate system of the Frame. The
Region class inherits from the Frame class, so a Region can be used in
place of the Frame that it contains.

The binding between the positions and the Frame is enforced by the Region
class through ``integrity restoration'' similar to that described in
section~\ref{sec:integrity}. If changes to the enclosed Frame are made
through a Region reference and they result in a change to the Frame's
coordinate system, then the coordinates stored in the Region are
automatically transformed into that new coordinate system.

In practice, the Region class is abstract. It does not have a constructor
function and is just a container class for other sub-classes which attach
particular semantics to the set of positions that they contain. The
simplest such sub-class is the PointList in which the positions are
simple independent points. For example, a PointList might contain a set
of star positions and a SkyFrame indicating that these are in ICRS
coordinates. If the SkyFrame's ``System'' attribute is changed via a Region
reference to represent FK5 coordinates, then the stored coordinates would
change so that the star positions remain fixed on the sky. As mentioned
above, however, the PointList is not much used for general coordinate
processing because of the overhead of packing and unpacking the object.

Other sub-classes of Region attach different semantics to the set of
positions enclosed. For example, there are sub-classes to represent
circles, ellipses, boxes, polygons, \emph{etc.}, where the enclosed positions
may be fixed in number and define the Region's shape (the centre and one
corner of a box, for example). These classes divide the Frame's domain
into an ``inside'' and an ``outside'' and, as the ``Region'' name suggests,
they are used to represent regions (intervals, areas, volumes, \emph{etc.})
within domains.

A negation method is provided that interchanges the inside and outside of
a Region. Separate Regions may also be combined in pairs using boolean
operations (AND, OR or XOR) and the CmpRegion class (which is analogous
to the CmpMap and CmpFrame classes described elsewhere). This allows
Regions of arbitrary complexity to be built out of simple components. A
Prism class is also provided that allows Regions to be extruded into
extra dimensions. For example, a 4-dimensional Prism can be formed by
extruding a 2-dimensional Circle and a 2-dimensional Box. A 4-dimensional
position is then regarded as inside the Prism if it lies inside the
Circle on axes 1 and 2 and inside the Box on axes 3 and 4.

The significance of the ``inside'' and ``outside'' of a Region lies in
its use as a Mapping, from which it also inherits. Typically, a Region
will map positions that lie inside it without change, but positions that
lie outside are mapped to a special null value. This provides a test of
whether a point is inside or outside a Region.

When combining Regions, it is not necessary for the coordinate systems
represented by their enclosed Frames to be the same, but it must be
possible to convert between them. If they differ, the \texttt{astConvert}
method
will be used to refer them both to a common coordinate system. Integrity
restoration also means that any changes to a Region's Frame that changes
the coordinate system will transform the boundary between the inside and
outside of the Region into the new coordinate system. This may alter the
nature of shapes such as circles, which might be distorted into ellipses
or even more elaborate shapes under non-linear coordinate
transformations. This is implemented by extending the enclosed Frame to
become a FrameSet, in which the original undistorted coordinate system is
preserved as the base Frame and the new coordinate system is the current
Frame. When testing whether a point lies within the Region, it is first
transformed from the current Frame into the base Frame of the Region's
FrameSet and then tested against the Region as defined in that coordinate
system.

In fact, Regions offer full support for enclosed FrameSets (as well as
Frames) because a FrameSet is a sub-class of Frame and can therefore be
supplied in place of a Frame when creating a Region. This allows a Region
to be defined that has a known shape in one coordinate system but can be
viewed through a range of other coordinate systems with corresponding
changes to its shape. This offers another means of creating Regions with
novel shapes.

One obvious application of Regions is in the segmentation of datasets -
for example, creating outlines that enclose objects or other regions of
interest in an image. However, by using Regions as Mappings and combining
them with other Mappings, it is possible to limit the range of
coordinates that the Mapping will accept, such that null values are
returned outside the Region. This effectively ``clips'' the range of
validity of the Mapping. This clipping can be arranged to occur in any
coordinate system: either the input or output space of the Mapping, an
intermediate space, or one devised purely for the purpose of clipping.
These techniques find particular use in graphical applications for
limiting the extent of drawing operations.

\section{Serialisation and FITS-WCS}
AST includes a set of \emph{Channel} classes, which allow AST objects to
be serialised in various ways for persistent external storage. The basic
Channel class has a method that converts an in-memory AST object into a
set of text strings. By default these text strings are simply written to
standard output, but an external ``sink'' function can be supplied to the
Channel constructor in order to redirect the text to some external data
store. The Channel class has another
method that does the inverse --- it calls a supplied ``source''
function to read a set of text strings from an external store, and then
creates an in-memory AST object from the text strings. A round-trip
(write followed read) is lossless. The textual descriptions of
AST objects produced by the Channel class use a bespoke block structured
format specific to AST.

As an example, the following Python code:

\begin{lstlisting}
map1 = Ast.PermMap([3,1],
                   [2,-1,1], 12.2)
map2 = Ast.ZoomMap(3, 4.0)
cmpmap = Ast.CmpMap(map1, map2)

channel = Ast.Channel()
channel.write(cmpmap)
\end{lstlisting}

produces the following output\footnote{Within the Python interface, the
\emph{print} function uses a Channel to produce a listing of the given
AST object. So the same output could more simply be produced by
``\texttt{print(cmpmap)}''.}:

\scriptsize
\begin{verbatim}
Begin CmpMap   # Compound Mapping
  Nin = 2     # Number of input coordinates
  Nout = 3    # Number of output coordinates
IsA Mapping    # Mapping between coordinate systems
  MapA =      # First component Mapping
     Begin PermMap    # Coordinate permutation
        Nin = 2       # Number of input coordinates
        Nout = 3      # Number of output coordinates
     IsA Mapping      # Mapping between coordinate systems
        Out1 = 2      # Output coordinate 1 = input coordinate 2
        Out2 = -1     # Output coordinate 2 = constant no. 1
        Out3 = 1      # Output coordinate 3 = input coordinate 1
        In1 = 3       # Input coordinate 1 = output coordinate 3
        In2 = 1       # Input coordinate 2 = output coordinate 1
        Nconst = 1    # Number of constants
        Con1 = 12.2   # Constant number 1
     End PermMap
  MapB =      # Second component Mapping
     Begin ZoomMap    # Zoom about the origin
        Nin = 3       # Number of input coordinates
     IsA Mapping      # Mapping between coordinate systems
        Zoom = 4      # Zoom factor
     End ZoomMap
End CmpMap
\end{verbatim}
\normalsize

Attributes of the Channel class can be used to enable or disable
inclusion of comments, indentation, defaulted values, \emph{etc}.

Sub-classes of Channel are provided that can encode this lossless
block-structured object description into an XML format (the XmlChan
class), or into a set of FITS header cards (the FitsChan class).

\subsection{Foreign Encodings}

In addition to the ``native'' block-structured object description outlined
above, AST can also read and write a range of ``foreign'' encodings that
allow data exchange with other systems. Because this is essentially a
format-conversion process it is not usually lossless. AST's flexibility
allows it to represent information that it may not be possible to store
in an external format, so some data loss may occur when writing and only
particular types of AST object can be successfully converted. Conversely,
when reading, there is normally little need to lose information, although
the more obscure features of some formats may not all be supported.

\subsubsection{FITS Encodings}
\label{sec:fitsencodings}
There are a number of conventions (unrelated to AST) for storing WCS
information in FITS header cards and the rules governing their use are
complex. Most programmers are unlikely to understand these technical
details, so AST wraps them up in a simple high-level model provided by
the FitsChan class. This provides a read/write interface between AST
objects and FITS header cards (where each card represents a
``keyword=value'' pair plus an optional comment).

The FitsChan class currently provides support for almost all of papers
I, II and III  in the FITS-WCS series \citep{FITSWCSI,FITSWCSII,FITSWCSIII}
\footnote{Support for the tabular format associated with the ``-TAB'' code
described in FITS-WCS paper III is limited to 1-dimensional (\emph{i.e.},
separable) axes.}. As yet it does not provide support for converting
between the AST TimeFrame class and the FITS-WCS description of time
axes \citep{FITSWCSIV}. It also supports a variety of unofficial encodings
that have been devised by particular projects over many years, including
several popular methods for describing focal plane distortion (such as
Spitzer SIP \citep{2005ASPC..347..491S} and SCAMP  TAN/TPV
\citep{2006ASPC..351..112B}).

Normally, FITS headers do not simply hold WCS information; they also
contain many additional cards describing other aspects of a dataset. Care
is therefore needed to ensure these other cards are not accidentally
removed or over-written while processing WCS information. Also, the
various FITS encodings of WCS information may overlap in their use of
FITS keywords, presenting an opportunity for keyword clashes.

To handle this, a FitsChan contains a buffer which may be loaded with
FITS header cards, normally read directly from a data file. The FitsChan
analyses these cards, determines how any contained WCS information has
been stored and makes this available through its ``Encoding'' attribute.
This takes values such as ``FITS-WCS'', ``FITS-IRAF'', ``NATIVE '',
\emph{etc.} to
indicate which convention has been used to store the information. A
subsequent call to the FitsChan's \texttt{astRead} method will then perform a
``destructive read'' on the buffer of FITS cards. This reads the AST object
that they describe and then deletes all the relevant FITS cards. The
method returns a FrameSet which holds the WCS information (see
section~\ref{sec:basecurrent}) and at the same time this information is
removed from the FITS header buffer, leaving it clear for new WCS
information to be written without the possibility of any keyword clashes.

To write a FrameSet\footnote{A FrameSet is the only valid object that
can be written to a FitsChan.} to a FITS header, a FitsChan's
\texttt{astWrite}
method is used. This uses the ``Encoding'' attribute to determine which
FITS header convention to use and then adds the resulting FITS header
cards to the FitsChan buffer, from where they may be transferred to an
output file. If changes have been made to a FrameSet such that it cannot
be stored using the chosen FITS encoding, then the \texttt{astWrite} method will
fail. AST will go to some lengths to simplify the information so that it
is compatible with the data model of the selected encoding, but that is
not always possible. In such cases, an alternative encoding can be chosen
simply by modifying the ``Encoding'' attribute of the FitsChan.

\subsubsection{The STC-S Encoding}
The StcsChan class provides an interface between AST and space-time
metadata from a subset of the ``STC-S'' scheme \citep{STC} developed by the
International Virtual Observatory Alliance (IVOA). This allows AST
Region objects to be converted to and from textual descriptions that
use the STC-S conventions, for example, converting the STC-S region
descriptions stored in catalogs created by the CUPID application
\citep[][\ascl{1311.007}]{2007ASPC..376..425B} and
displaying the regions over images displayed in Starlink GAIA
\citep{2010ASPC..434..213B}.

\section{Fields of Application}
Whilst the AST library provides many facilities that are useful when
describing and using the WCS information attached to a data array, it
is not limited to that field. Its generalised design enables it to be used
in any situation where relationships between several different coordinate
systems need to be managed. Just to emphasise that point, WCS handling is
included as the last item in this section.

\subsection{Generalised Plotting}
\label{sec:plotting}
Most graphics systems provide many coordinate systems through which the
display can be addressed - device pixel coordinates, normalised viewport
coordinates, \emph{etc}. However, since the details of the relationships
between such coordinate systems may vary from system to system, AST does
not attempt to describe them all. Instead it just assumes there is at
least one such coordinate system available and provides facilities that
allow it to be attached to that system and to use it to perform
plotting\footnote{The user is still free to use the underlying graphics
system directly when required.}. In addition, AST allows application
software to define extra --- potentially non-linear --- coordinate
systems, and to integrate them with the coordinate system provided by the
underlying graphics package. For this purpose AST provides two classes ---
\emph{Plot}, which provides two-dimensional plotting facilities, and
\emph{Plot3D}, which provides three-dimensional plotting facilities.

\subsubsection{Use of External Plotting Packages}
The plotting classes within AST do not themselves include any facilities
for placing ``ink on to paper'' --- an external graphic library must be
supplied to draw graphical primitives such as straight lines, markers and
character strings. The plotting classes within AST will then use the
primitive facilities of this underlying graphics system to draw more complex
entities such as annotated coordinate grids, \emph{etc}.

This separation between coordinate handling and drawing enables the
sophisticated plotting capabilities of AST to be used within many
different systems and languages\footnote{It is known that AST graphics
have been used with plotting packages written in Java, Perl, Tk-Tcl and
Python, as well as C and Fortran.}.

Each of the underlying graphics systems (the 2-d and 3-d plotting classes use
separate systems) can be specified in two ways:

\begin{enumerate}
\item At build-time. In this method, a module must be supplied (callable from
C) that provides implementations of a set of wrapper functions that AST
uses to perform primitive drawing operations. Each such wrapper function
makes appropriate calls to the underlying graphics system to perform its
work. This module is linked into the executable at build-time, in place
of the default module provided by AST.
\item At run-time. In this method, application code registers pointers to
the graphics wrapper functions with AST, whilst the executable is running.
The registered function pointers are used in preference to any functions
specified at build-time.
\end{enumerate}

The AST library includes modules that allow the PGPLOT
graphics package \citep[][\ascl{1103.002}]{1991BAAS...23..991P} to
be used for drawing by both the Plot class and the Plot3D class (see the
following sections).

\subsubsection{Two Dimensional Plotting}
A \emph{Plot} is a subclass of FrameSet and can therefore be considered
to be a FrameSet ``with some extra facilities''. A Plot does not represent
the graphical content itself, but is a route through which plotting
operations, such as drawing lines and curves, are conveyed on to a
plotting surface to appear as visible graphics.

When considered as a FrameSet, the base Frame within a Plot corresponds
to the Cartesian coordinate system used to specify positions to the
underlying graphics system\footnote{This coordinate system must have been
defined previously using appropriate calls to the underlying graphics
system.}. The bounds of the plotting area within this coordinate system must
be specified when the Plot is created. A typical Plot may for instance
have a base Frame that describes millimetres from the bottom left corner
of the plotting area.

The current Frame within a Plot corresponds to ``user''  coordinates ---
\emph{i.e.}, the coordinate system in which the application code wishes to
specify positions. Since the Plot is a form of FrameSet, it will also
include the Mappings needed to transform ``user'' coordinates into the
corresponding graphics coordinates. A typical Plot may for instance
have a current Frame that describes (RA,Dec) positions on the sky.

When a Plot is created, an existing FrameSet is supplied together with
the bounds of a box within its base Frame (typically image pixel
coordinates). A Mapping (either linear or logarithmic) is created that
maps this box on to a specified area within the graphics coordinate
system. The Plot constructor then initialises the new Plot to be a copy
of the supplied FrameSet, and adds in a new Frame describing the graphics
coordinates system, using the above Mapping to connect the new Frame to
the base Frame. This new Frame is then made the base Frame in the Plot.
In effect, the Plot becomes attached to the plotting surface in rather
the same way that a basic FrameSet might be attached to (say) an image.

The Plot class has methods that can draw markers, geodesic curves, text
strings, \emph{etc}. The application code supplies the positions of these
objects within the ``user'' coordinate system (\emph{i.e.}, the current
Frame of the Plot --- \emph{e.g.}, (RA,Dec)). The Plot class then uses the
Mappings stored within the Plot to transform them into the graphics
coordinate system (\emph{i.e.}, the base Frame of the Plot), before
invoking the appropriate wrapper functions to instruct the underlying
graphics system to draw the required primitives. Whilst this is a fairly
simple process for items such as graphical markers that only have one
associated coordinate (\emph{i.e.}, the centre of the marker), it is a
more complex process for items such as geodesic curves that span a wide
range of different coordinates. In this case, the curve is represented by
a set of positions spread along the curve in user coordinates. All these
positions are transformed into graphics coordinates before being plotted
in the form of a ``poly-line''. The density of points varies along this
line, and is chosen to ensure that any discontinuities or highly
non-linear sections of the curve are drawn with sufficient accuracy. All
aspects of the generated plot can be controlled via attributes of the Plot
class.

\begin{figure*}[ht]
\centering
\includegraphics[width=\textwidth]{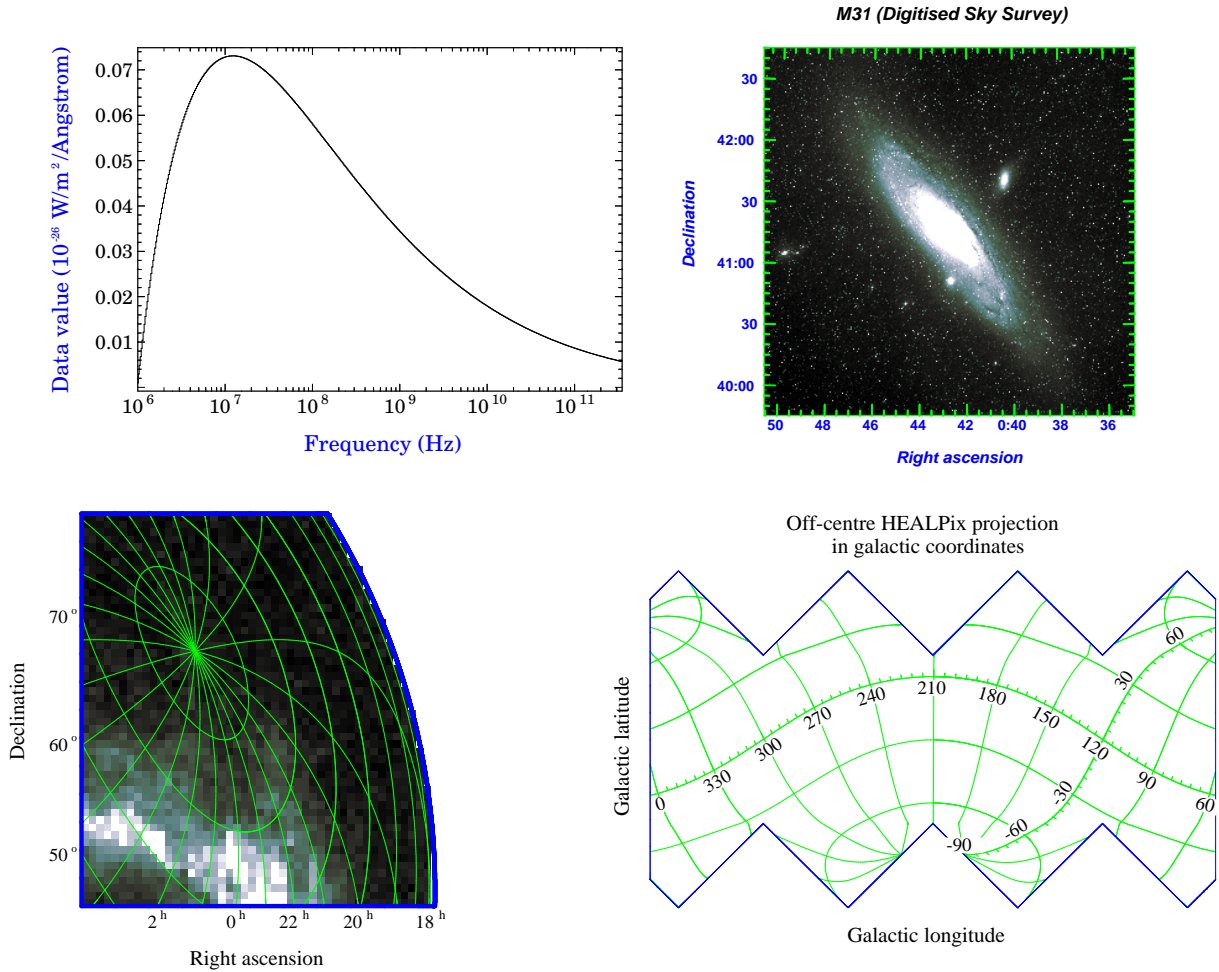}
\caption{A selection of annotated axes created by the Plot class. Note,
Plot only creates the annotated axes --- the background images were
produced using PGPLOT directly. The PGPLOT library was used as the
underlying graphics package for these plots.}
\label{fig:2dplots}
\end{figure*}

The most comprehensive drawing method supplied by the Plot class produces
a complete set of annotated axes describing the area of user coordinates
visible within the plotting area. Some examples are shown in
Fig.~\ref{fig:2dplots}.

\subsubsection{Three Dimensional Plotting}

The \emph{Plot3D} class extends the Plot class to provide plotting
facilities in three dimensions. The basic model is the same as for the
Plot class --- a Plot3D object is a FrameSet in which the base Frame
represents the 3-dimensional Cartesian coordinate system used by the
underlying 3-dimensional plotting package, and the current Frame
represents 3-dimensional ``user'' coordinates.

The projection from 3-dimensional graphics coordinates on to a
2-dimensional plotting surface is handled by the underlying graphics
package. This may be surprising, given that AST could itself handle this
sort of projection. But the decision was taken so that Plot3D graphics
wrapper functions could use the full range of features offered by a
dedicated 3-dimensional plotting package, rather than restrict it to the
more limited features offered by a typical 2-dimensional package.

\begin{figure}[h]
\centering
\includegraphics[width=\columnwidth]{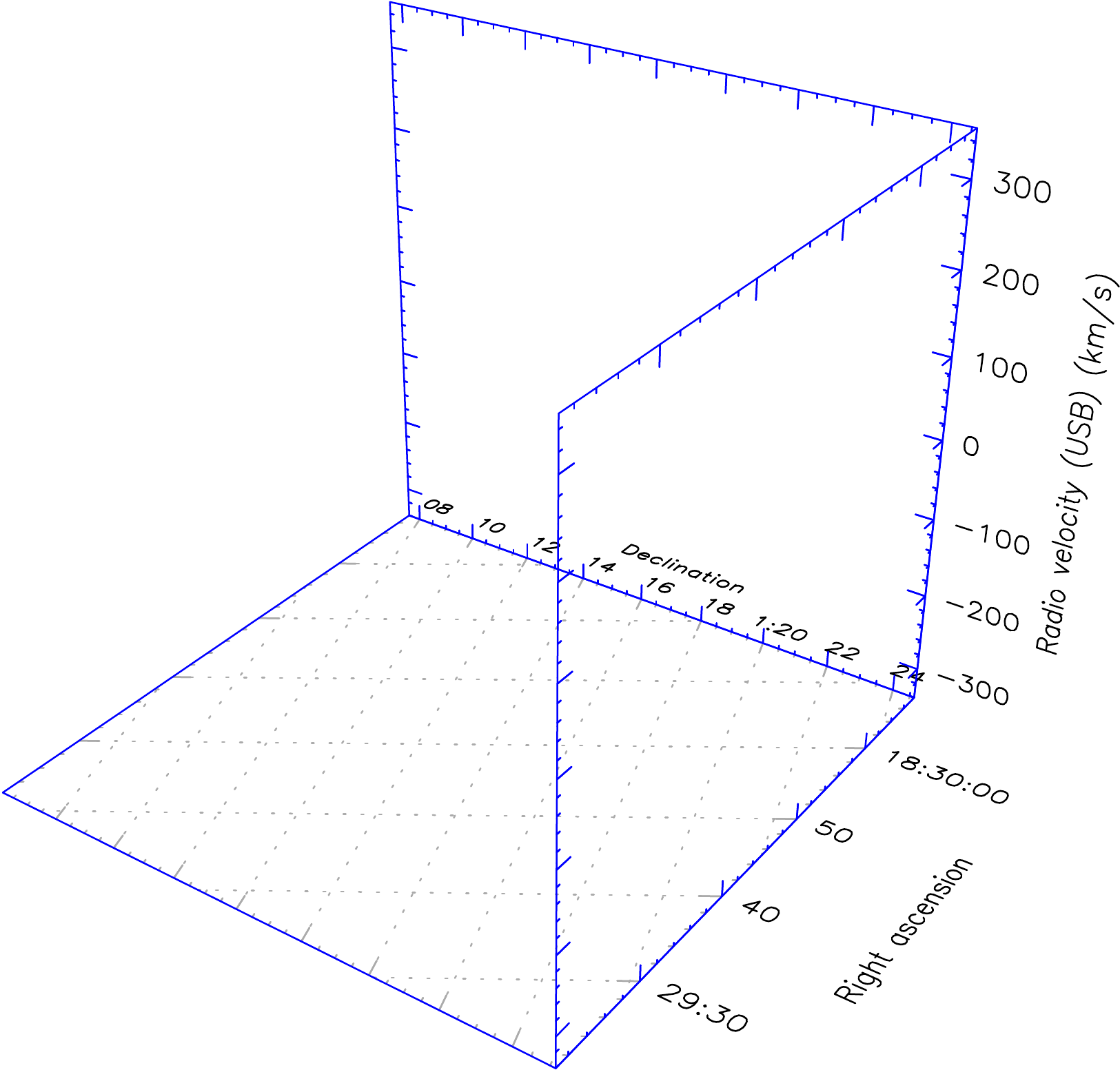}
\caption{Annotated axes created by the Plot3D class. The 3D interface to the
PGPLOT library included with AST was used as the underlying graphics package
for these plots.}
\label{fig:3dplot}
\end{figure}

Fig.~\ref{fig:3dplot} shows an example of the 3-dimensional annotated axes
produced by Plot3D. The Plot3D class was originally developed to
support 3-D coordinate grid plotting in the Starlink GAIA display tool
\citep[][\ascl{1403.024}]{2008ASPC..394..339D} using the VTK 3-d graphics
package \citep{Hanwell2015}.

\subsection{Flux and Data Unit Transformation}

In the majority of scientific data sets, each data point has a
\emph{position} and a \emph{value}. Whilst the concepts described in this
paper may seem more naturally associated with the positional information,
they can also be applied to the value (or values) associated with each
data point. A position within some N-dimensional space is specified by a
set of N axis values. Likewise, a value can also be represented by a set
of axis values. In the majority of common cases data values are
1-dimensional - for instance a temperature, or a sky brightness. But
there are also some common multi-dimensional cases such as a Stokes vector
$(I,~Q,~U,~V)$.

The functional division into Mappings, Frames and FrameSets  used
throughout AST is equally applicable to the problem of describing data
\emph{values} as it is to describing data \emph{positions}. To do so
requires a set of specialised Frame classes to be created to describe each
set of equivalent data value systems. For instance a hypothetical
\emph{TemperatureFrame} could be created that encapsulates the meta-data
needed to transform between different temperature scales (Celsius, Fahrenheit,
Kelvin, \emph{etc}).

Currently, provision of such specialised Frames within AST is
limited, but this may change in future:

\begin{description}

\item[FluxFrame :] Supports 1-dimensional axes that represent the following
astronomical flux systems:
\begin{itemize}
\item Flux per unit frequency ($W/m^2/Hz$)
\item Flux per unit wavelength ($W/m^2/Angstrom$)
\item Surface brightness in frequency units ($W/m^2/Hz/arcmin^2$)
\item Surface brightness in wavelength units ($W/m^2/Angstrom/arcmin^2$)
\end{itemize}
Any dimensionally equivalent units can be used in place of the default
units listed above. All flux values are assumed to be measured at the
same frequency or wavelength (specified by an attribute of the FluxFrame).
Thus this class is more appropriate for use with images than spectra.

\item[Frame :] The basic Frame class can be used to represent
single-system data value axes (\emph{i.e.}, Frames that know only one
system for specifying positions within its domain), using any suitable
system of units. Such Frames can be used to convert axis values into any
dimensionally equivalent set of units. For instance, a basic Frame used
to represent stellar mass could be set to use kg, Solar mass,
\emph{etc.} as its units, and will automatically modify the associated
Mappings each time the units are changed (assuming the Frame is part of
a FrameSet).

\end{description}

\subsection{Attaching WCS Information to Datasets}
The most common reason for using AST is to allow the position of one or
more data values to be described in a range of possible coordinate
systems. For this purpose, a serialised FrameSet is often stored with the
data on the assumption that the ``natural coordinate system'' of the data
structure corresponds to a known Frame (usually the base Frame) in the
FrameSet. The most common data structure is an N-dimensional regular grid
of values, in which case the base Frame in the FrameSet is assumed to describe
``pixel coordinates'' within the grid (generalising the term ``pixel'' to
data of any dimensionality). Different systems exist for enumerating the
pixels within an array - for instance, some start counting at zero and
some at one. AST allows multiple pixel coordinate systems to be described
and used within a single FrameSet, assuming that the Mappings between
such Frames have been set up and incorporated into the FrameSet correctly.

The following sections describe just a few of the more common WCS
operations that may be achieved using AST.

\subsubsection{Validating WCS}
The generalised description of WCS provided by AST makes it possible
to write applications in a domain-agnostic manner. For instance, an
application that uses WCS to align two data sets need not know whether the
data sets are two-dimensional images of the sky, one dimensional spectra,
3-dimensional time-space cubes, \emph{etc}. However, this will not always
be possible, and it will still often be the case that an application needs
to check that the WCS in the supplied data is consistent with the
specific requirements of the application. The \texttt{astFindFrame} method can
often be used for this purpose. The application creates a Frame that acts
as a template for the coordinate systems required by the application,
and passes this Frame, together with the WCS FrameSet read from the data,
to the \texttt{astFindFrame} method, which then searches the FrameSet looking for
a Frame that can be matched against the template. If such a Frame is
found, information about the specific Frame found is returned, together
with a Mapping that connects that Frame to one of the Frames in the WCS
FrameSet.

As a trivial example, consider an application that requires 1-dimensional
data but does not care what that one dimension represents. It creates a
1-dimensional basic Frame to act as a template, leaving all Frame
attributes at their default values in the template (default values act as
wild-cards during the searching process). The application then passes
this template Frame, together with the WCS FrameSet read from the data
file, to \texttt{astFindFrame}. Each Frame in the WCS FrameSet is then compared to
the template to see if a match is possible. In this case, a basic Frame
with no set attributes will match any 1-dimensional Frame of any form,
but will not match Frames with more than one dimension.

A specialised application for processing spectral data cubes may use for
its template a 3-dimensional CmpFrame (compound Frame) containing a
2-dimensional SkyFrame (celestial longitude/latitude axes) and
1-dimensional SpecFrame (a spectral axis). The \texttt{astFindFrame} method will then
only match 3-dimensional Frames with similar properties. Note, if the
System attribute is set to specific values in the SkyFrame and/or
SpecFrame contained within the template, then the Mapping returned by
\texttt{astFindFrame} will include the transformations needed to convert from the
system of the matching Frame to that of the template Frame. So for
instance, if the template axes are $(RA,Dec,Wavelength)$ and the WCS
FrameSet contains a CmpFrame with $(Frequency,GalacticLongitude,
GalacticLatitude)$ axes, the Mapping returned by \texttt{astFindFrame} will include an
axis permutation, and the transformations needed to convert from frequency
to wavelength, and from Galactic coordinates to $(RA,Dec)$.

In other words, \texttt{astFindFrame} provides arbitration between the WCS
requirements of the application, and the WCS information that is available
in the supplied data set, returning a null result if the requirements of
the application are not met by the data set.

\subsubsection{Merging WCS information (Alignment)}
If a data processing operation involves combining two or more datasets in
some way, it will usually be necessary to form a connection between the
WCS in the two datasets, so that corresponding elements in each dataset
can be identified. This process can be considered one of alignment. For
example, in the case of images of the sky, we may wish to align them
prior to combining them into a mosaic. If each dataset is calibrated
using a FrameSet whose base Frame corresponds with the underlying data
elements (\emph{e.g.}, pixels), then alignment involves finding a transformation
between the base Frames of the two FrameSets involved.

This operation is performed using the \texttt{astConvert} method introduced in
section~\ref{sec:domConversion}, whose role is to find a transformation
between the coordinate systems represented by two Frames. Because
FrameSets are also Frames, \texttt{astConvert} may be used to convert between two
FrameSets. In this case, however, there is a choice of how to find a
transformation between the two Frames that they represent (their current
Frames). This is because links between two FrameSets can potentially be
formed in a number of places by matching Frames within each FrameSet that
have the same domain.

An example should make this clear. Suppose we have two images, each
calibrated with FrameSets whose base Frames represent image pixel
coordinates. Also suppose that each FrameSet contains two additional
Frames that give the image's position in the focal plane of a telescope
(mm) and its position on the sky (RA,Dec). If we want to align these
images, we can do it in three ways by matching: (1) corresponding pixels,
(2) corresponding focal plane positions or (3) corresponding positions on
the sky. Any of these is a legitimate way to align the two images and may
be appropriate for different purposes.

When presented with the two FrameSets,\footnote{Because astConvert acts
on the current Frame of a FrameSet, the two FrameSets must actually be
temporarily modified to interchange their base and current Frames.
Inverting them will have this effect.} astConvert has the same
choices. This ambiguity may be removed by supplying the name of the
domain that we wish to use for alignment. In this example, we might
select ``SKY'' and astConvert would then try to find a route between the
two FrameSets that joins a pair of SkyFrames (representing celestial
coordinates), one from each FrameSet. A domain ``search path'' may also be
given, in which case each listed domain will be tried in turn until one
succeeds.

The result returned by astConvert will be a FrameSet which, when used as
a Mapping, converts pixel coordinates in one image into pixel coordinates
in the second image, such that the two images are aligned in the selected
domain.

\subsubsection{Modifying WCS Information}

If an application creates an output data array by applying some geometrical
transformation to an input data array, it should also store appropriately
modified WCS in the output. The concatenation and simplification of
Mappings described in section~\ref{sec:mappings} makes this easy. The
application should create a Mapping that describes the geometric
transformation that has been applied to the pixel array, and then
``re-map'' the pixel Frame within the WCS FrameSet read from the input
data set. The resulting modified FrameSet should be stored in the output
data set. Re-mapping a Frame within a FrameSet means appending the
supplied Mapping to the existing Mapping that connects the Frame with its
parent in the FrameSet. A method is provided to do this, which also simplifies
the resulting compound Mapping if possible.

For instance, if an application rotates an input image to create an
output image, the application should create a 2-dimensional MatrixMap to
describe the rotation\footnote{A pair of ShiftMaps will also be needed if
the rotation is not around the pixel origin.} and then invoke
\texttt{astRemapFrame}, supplying the MatrixMap and the WCS FrameSet from the
input image. The modified FrameSet would then be stored in the output
image.

\subsubsection{Using WCS for Data Resampling and Regridding}
The previous section described how to modify the WCS to take account of a
geometric transformation of a data set, assuming a Mapping describing the
transformation is available. If such a Mapping is available, it can also
be used to perform the actual resampling or regridding of the pixel
values themselves. AST provides several methods that will perform such
resampling or regridding, using any of a wide range of alternative
sampling kernels. Alternatively, an externally defined sampling kernel
can be used.

The benefit of using these methods within AST rather than simply
transforming every pixel position within the application code, is that
the AST methods attempt to speed up the operation by using linear
approximations to the supplied transformation if possible. Data sets are
constantly increasing in size, and transforming every pixel position
within a large data set using a long and complicated transformation can
be an expensive operation. The AST methods divide the input data set in
half along each axis, and find a linear approximation to the Mapping
over each resulting quadrant\footnote{The word ``quadrant'' is used here
for clarity but does not imply a restriction to two dimensions. The
algorithm itself can operate in any number of dimensions.}. If one of
these approximations meets a
user-specified accuracy, the approximation is used to transform all pixel positions
within its quadrant. Any quadrants that do not meet the accuracy
requirements are subdivided again and new approximations are found
for each of the new sub-quadrants. This process repeats recursively until
the sub-quadrants become so small that there is likely to be little
time-saving in sub-dividing them any further. At this point the full
transformation is used on all pixels within any such sub-quadrants.

\section{Things we Would do Differently Now}

\subsection{Language Choice}

There was no C++ standard in 1996 when AST development commenced.
So for portability reasons we chose to write AST in ANSI C, developing
our own infrastructure to support object-orientation. Things are
different now --- C++ is much more mature with greater standardisation.
For that reason we would choose C++ over C if starting AST development
now\footnote{Writing in higher level languages such as Java or Python would make
it more difficult to write interfaces for other languages. Conversely,
lack of native Java implementation was a significant impediment to
adoption for applications desiring true platform independence such as TOPCAT
\citep[][\ascl{1101.010}]{2005ASPC..347...29T}.}

The distinction between C and C++, of course, only affects the way that the
class interfaces are implemented and does not impact on the AST data model
nor, indeed, on the majority of the AST code. Consequently, converting AST
to use C++ would not necessarily involve a major re-write.

However, using C as an implementation language did necessitate the AST
public interface being wrapped in a set of macros. These were not used
internally within the code and this resulted in a distinction between the
public interface and the internals of AST, with clear differences in the
way they are called. As a consequence, users of the public interface have
found accessing the internals of AST to be challenging and this has made
extending the system more difficult. Using C++ would have mitigated this
as the language provides mechanisms for protecting internal interfaces in
a more controlled manner. However, some of the features of the existing
AST public interface - such as pointer validity checking and object
scoping - could potentially be difficult to provide in C++.

\subsection{Architecture}

Packing the wide functionality of AST into a single monolithic
library and a large document has sometimes proved to be a barrier to
adoption. Potential users may want to use only a small part of the
functionality available and feel daunted by a system that addresses
many problems that are irrelevant to them.

A more open and modular architecture consisting of a set of optional
components together with a small set of of mandatory libraries
providing the core functionality would probably have had wider appeal.
For instance, the plotting and region classes within AST would have
made good candidates for optional extensions.

Such an architecture could also make it easier for extensions to be
written by people and groups not directly involved with the
development of the core functionality of AST\footnote{Writing AST in C++
rather than C would also have made this easier, as noted in the previous
section.}.

The problem remains, however, that persistent objects written by
extensions to AST could not be handled by other AST installations that
lack the necessary extra classes. This could seriously affect data
portability. A framework that allowed AST to be extended by means of
"plug-in" modules might be a solution, so that unrecognised data
resulted in a message asking the user to install the required extra
software. However, this adds considerable complexity and presents a
further barrier to extending the system because developers must learn
how to build the plug-in modules.\footnote{This is similar to the
  problems that the HDF5 community have to consider when handling
  data compression filters \citep{Folk:2011:OHT:1966895.1966900}.}

This tension between extensibility and the portability of persistent
data seems to be an inherent feature of OO design, despite its many
other advantages. Currently, we are not aware of a good lightweight
solution to the problem.

\subsection{Floating-point precision}

The restriction that all coordinate values are represented by
double precision values may be a problem for time axes, which can
sometimes require quad precision.

\subsection{Coordinate systems versus Domains}

The distinction between a coordinate system and a domain has caused
confusion and misunderstandings on several occasions. Coordinate systems are
mathematical abstractions of various types (Cartesian, polar, \emph{etc}).
Coordinate systems are used to describe positions within a physical space
(\emph{i.e.}, a domain). In this sense a domain can encapsulate several
alternative coordinate systems, any of which can be used to describe positions
in the domain. With hind-sight, it may have been useful to make this
distinction clearer by separating coordinate systems and domains into
different classes.

\subsection{Angle units}

Our decision to use radians rather than degrees to represent angle on the
sky has proved unpopular. The ubiquity of the FITS data format within the
astronomical community has led to an expectation that other systems will
follow the example of FITS and use degrees rather than radians to
represent spherical coordinates. Failure to check the AST documentation on
this point has been a common source of bugs within software that uses AST.

Normally, allowing the use of different units for angles would be
straightforward given AST's ability to convert between equivalent units.
However, the SkyFrame class was implemented long before unit conversion
was added to the library and in a manner that conflated unit conversion
with formatting (through the use of format strings such as
``\texttt{hh:mm:ss}'').
This, unfortunately, leaves the SkyFrame class unable to benefit from
automatic unit conversion. This is an area of some subtlety that future
designs would do well to consider carefully.

\section{The AST library}

The AST library has been developed over a number of years
\citep{1998ASPC..145...41W,2000ASPC..216..506W,2001ASPC..238..129B,2004ASPC..314..412B,2008ASPC..394..635B,2010ASPC..434..213B,2012ASPC..461..825B}
  and is written in C with no dependencies. It includes code from
  WCSLIB \citep[][\ascl{1108.003}]{2006ASPC..351..591C}, PAL \citep{2013ASPC..475..307J}
  and SOFA \citep[][\ascl{1403.026}]{2011SchpJ...611404H} but does not depend on those
  libraries. There are language bindings for Fortran, Java, Perl and
  Python. The
source code for the AST library is open-source using the GNU Lesser General
Public License and is available
online\footnote{\url{https://github.com/Starlink/starlink}}.

\section{Acknowledgements}

This work was funded by the Council for the Central Laboratory of the
Research Councils and subsequently by the Science and Technology
Facilities Council.  The AST library and the related Starlink software
are currently maintained by the East Asian Observatory, Hawaii.

\bibliographystyle{model2-names-astronomy}
\bibliography{acast}







\end{document}